\documentclass[preprint,11pt,nofootinbib]{revtex4-1}

\usepackage{graphicx}
\usepackage{dcolumn} 
\usepackage{bm}
\usepackage{amssymb}
\usepackage{amsmath}
\usepackage{epsfig}    
\usepackage{color}
\usepackage{slashed}
\usepackage{hyperref}
\begin{document} 

\title{Implication of the $W$ boson  mass anomaly at CDF II\\ in the Higgs triplet model with a mass difference}
\preprint{OU-HET-1141}
%\date{\today}

\author{Shinya Kanemura}
\email{kanemu@het.phys.sci.osaka-u.ac.jp}
\affiliation{Department of Physics, Osaka University, Toyonaka, Osaka 560-0043, Japan}

\author{Kei Yagyu}
\email{yagyu@het.phys.sci.osaka-u.ac.jp}
\affiliation{Department of Physics, Osaka University, Toyonaka, Osaka 560-0043, Japan}

\begin{abstract}

A new report for the measurement of the W boson mass has been provided by the CDF II experiment. 
The measured value of the W boson mass is given to be $m_W^{\text{CDF II}} = 80.4335 \pm 0.0094$ GeV which shows $7\sigma$ deviation from the standard model prediction, while 
the other groups of high energy experiments, e.g., ATLAS and D0 II, show consistent results with the latter. 
Assuming that the measurement at CDF II is correct, we discuss the possibility to explain the anomaly in the Higgs triplet model. 
We find that the anomaly can be explained under the constraints from the electroweak rho parameter, the effective weak mixing angle, the partial width of the leptonic $Z$ decay and current direct searches at LHC when 
non-zero mass differences among the triplet-like Higgs bosons and the vacuum expectation value of the triplet Higgs field are 
taken to be ${\cal O}(100)$ GeV and ${\cal O}(1)$ GeV, respectively.

\end{abstract}
\maketitle

%%%%%%%%%%%%%%%%%%%%%%%%%%%%%%%%%%%%%%%%%%%%%%%%%%
%\section{Introduction}
\noindent
{\it Introduction ---}
%%%%%%%%%%%%%%%%%%%%%%%%%%%%%%%%%%%%%%%%%%%%%%%%%%
%
Nearly a decade after the discovery of the Higgs boson at LHC, the true shape of the Higgs sector is still unknown. 
On the other hand, the Higgs sector is often extended from the minimal form in the standard model (SM) for
models beyond the SM (BSM), which can explain neutrino oscillations, dark matter and baryon asymmetry of the Universe. 
Therefore, unveiling the structure of the Higgs sector is quite important to narrow down BSM scenarios.

Recently, the CDF II collaboration has reported new results for the measurement of the W boson mass using the data set with the integrated luminosity of 8.8 fb$^{-1}$. 
Surprisingly, the observed value, $m_W^{\text{CDF II}} = 80.4335 \pm 0.0094$ GeV, deviates $7\sigma$ level from the SM prediction, i.e., $m_W^{\rm SM} = 80.357\pm 0.006$ GeV~\cite{CDF:2022hxs}. 
On the other hand, this result gives rise to a tension with that given by the other groups, e.g., ATLAS and D0 II, which show the consistent result with the SM prediction. 
Although we need to understand the origin of such a tension, it is worth to discuss possible implications of the large deviation in the W boson mass assuming the result by CDF II being correct. 
In fact, just a few days after the announcement, a number of papers have already appeared~\cite{Arias-Aragon:2022ats,DiLuzio:2022xns,Lu:2022bgw,Babu:2022pdn,Han:2022juu,Athron:2022isz,Athron:2022qpo,Ghoshal:2022vzo,Ahn:2022xeq,Heo:2022dey,Krasnikov:2022xsi,Gu:2022htv,Asadi:2022xiy,Sakurai:2022hwh,Blennow:2022yfm,Strumia:2022qkt,Du:2022pbp,Fan:2022dck,Bahl:2022xzi,Endo:2022kiw,Kawamura:2022uft,Crivellin:2022fdf} in the context that the W boson anomaly is a signature of BSM scenarios. 

In this Letter, we discuss implication of the W boson mass anomaly in models with extended Higgs sectors because the latter can strongly be related to various BSM scenarios as mentioned above. 
We particularly focus on the Higgs triplet model (HTM) which appears in the type-II seesaw model~\cite{Cheng:1980qt,Schechter:1980gr,Lazarides:1980nt,Mohapatra:1980yp}, 
where the Higgs sector is composed of an isospin doublet $\Phi$ with the hypercharge $Y = 1/2$ and an isospin triplet $\Delta$ with $Y = 1$. 
One of the quite striking features of this model is that the electroweak (EW) rho parameter deviates from unity at tree level 
due to non-zero values of the vacuum expectation value (VEV) of the triplet field. 
This means that four, not three as in the SM, input parameters are required to describe the EW sector. 
We take the fine structure constant $\alpha_{\rm em}$, the Fermi constant $G_F$, the Z boson mass $m_Z^{}$ and the triplet VEV $v_\Delta$ as the four input parameters, 
and then calculate the W boson mass $m_W$, the EW rho parameter, the effective weak mixing angle and the partial width of the leptonic Z decay at one-loop level based on the renormalization scheme developed in Refs.~\cite{Blank:1997qa,Kanemura:2012rs,Aoki:2012jj}. 
We will see below that there are parameter regions that can explain the anomaly under the constraints from the measurements of the above EW parameters and the current direct searches at LHC. 

%%%%%%%%%%%%%%%%%%%%%%%%%%%%%%%%%%%%%%%%%%%%%%%%%%
%\section{Model \label{sec:model}}
\noindent
{\it Model ---}
%%%%%%%%%%%%%%%%%%%%%%%%%%%%%%%%%%%%%%%%%%%%%%%%%%
We parameterize the doublet $\Phi$ and the triplet $\Delta$ fields as 
\begin{align}
\Phi=\left[
\begin{array}{c}
\phi^+\\
\frac{1}{\sqrt{2}}(\phi+v_\phi+i\chi)
\end{array}\right],\quad \Delta =
\left[
\begin{array}{cc}
\frac{\Delta^+}{\sqrt{2}} & \Delta^{++}\\
\Delta^0 & -\frac{\Delta^+}{\sqrt{2}} 
\end{array}\right]\text{ with } \Delta^0=\frac{1}{\sqrt{2}}(\delta+v_\Delta+i\eta), \label{eq:field}
\end{align}
where $v_\phi$ and $v_\Delta$ are the VEVs of $\Phi$ and $\Delta$, respectively which satisfy 
$v^2\equiv v_\phi^2+2v_\Delta^2 = (\sqrt{2}G_F)^{-1}\simeq$ (246 GeV)$^2$. In this model, the masses of the $W$ and $Z$ bosons are obtained at tree level as
\begin{align}
m_W^2 = \frac{g^2}{4}v^2,\quad m_Z^2 =\frac{g^2}{4c^2_W}(v^2+2v_\Delta^2),  \label{mV}
\end{align}
where $c_W^{} = \cos\theta_W$ and $s_W^{} = \sin\theta_W$ with $\theta_W$ being the weak mixing angle. 
The EW rho parameter thus deviates from unity at tree level:
\begin{align}
\rho_{\rm tree} \equiv \frac{m_W^2}{m_Z^2c^2_W}=\frac{v^2}{v^2+2v_\Delta^2}. \label{rho_triplet}
\end{align}
From EW global fits, the rho parameter is given to be $\rho = 1.0002\pm 0.0009$~\cite{Zyla:2020zbs}, so that the
value of $v_\Delta$ is constrained to be $v_{\Delta} \lesssim  7$ GeV at 95\% confidence level (CL) using the tree level formula. 

The most general Higgs potential is given by 
\begin{align}
V(\Phi,\Delta)&=m^2\Phi^\dagger\Phi+M^2\text{Tr}(\Delta^\dagger\Delta)+\left[\mu \Phi^Ti\tau_2\Delta^\dagger \Phi+\text{h.c.}\right]\notag\\
&+\lambda_1(\Phi^\dagger\Phi)^2+\lambda_2\left[\text{Tr}(\Delta^\dagger\Delta)\right]^2+\lambda_3\text{Tr}[(\Delta^\dagger\Delta)^2]
+\lambda_4(\Phi^\dagger\Phi)\text{Tr}(\Delta^\dagger\Delta)+\lambda_5\Phi^\dagger\Delta\Delta^\dagger\Phi, \label{pot_htm}
\end{align}
where all the parameters are taken to be real without loss of generality. 
The relation between the weak basis defined in Eq.~(\ref{eq:field}) and the mass basis of the Higgs bosons is given by 
\begin{align}
\Delta^{\pm\pm}=H^{\pm\pm},~~
\left(
\begin{array}{c}
\phi^\pm\\
\Delta^\pm
\end{array}\right)&=
R(\beta)
\left(
\begin{array}{c}
G^\pm\\
H^\pm
\end{array}\right),~ 
\left(
\begin{array}{c}
\chi\\
\eta
\end{array}\right)=
R(\beta')
\left(
\begin{array}{c}
G^0\\
A
\end{array}\right),~
\left(
\begin{array}{c}
\phi\\
\delta
\end{array}\right)=
R(\alpha)
\left(
\begin{array}{c}
h\\
H
\end{array}\right),
%~\text{with}~
%R(\theta) = \begin{pmatrix}
%\cos\theta & -\sin\theta \\
%\sin\theta & \cos\theta
%\end{pmatrix}
  \label{mixing1}
\end{align}
with $R(\theta)$ being the $2\times 2$ orthogonal matrix. 
In the above expression, $G^\pm$ and $G^0$ are the Nambu-Goldstone bosons absorbed into the longitudinal component of the W and Z bosons, respectively. 
The mixing angles are given by 
\begin{align}
\tan\beta=\frac{\sqrt{2}v_\Delta}{v_\phi},\quad \tan\beta' = \frac{2v_\Delta}{v_\phi}, \quad
\tan2\alpha &=\frac{v_\Delta}{v_\phi}\frac{2v_\phi^2(\lambda_4+\lambda_5)-4M_\Delta^2}{2v_\phi^2\lambda_1-M_\Delta^2-2v_\Delta^2(\lambda_2+\lambda_3)}, \label{tan2a}
\end{align}
where $M_\Delta^2\equiv v_\phi^2\mu/(\sqrt{2}v_\Delta)$. 
We see that in the limit of $v_\Delta/v_\phi\to 0$, $H^{\pm\pm}$, $H^{\pm}$, $A$ and $H$ are purely the Higgs bosons coming from $\Delta$, 
so that we call them as triplet-like Higgs bosons. On the other hand, couplings of $h$ take the same values as those of the SM Higgs boson at tree level. We thus can identify $h$ with the discovered Higgs boson with a mass of 125 GeV. 
In this limit, we obtain the following relation among the masses of the triplet-like Higgs bosons
\begin{align}
m_{H^{\pm\pm}}^2-m_{H^{\pm}}^2&=m_{H^\pm}^2-m_{A}^2~~\left(=-\frac{\lambda_5}{4}v^2\right), \quad m_A^2=m_{H}^2~~(=M_\Delta^2). 
\end{align}
Thus, the mass spectrum is determined by the two parameters, i.e., 
the mass of the lightest triplet-like Higgs boson $m_{L}^{}$ and the squared mass difference $\Delta m^2 \equiv -\lambda_5v^2/4$. 
We note that depending on the sign of $\Delta m^2$ there are two patterns of the mass hierarchy, i.e., $m_{H^{\pm\pm}} > m_{H^{\pm}} > m_{H}$ for $\Delta m^2 > 0$ and 
$m_{H^{\pm\pm}} < m_{H^{\pm}} < m_{H}$ for $\Delta m^2 < 0$. 

Under $v_\Delta/v_\phi \ll 1$, productions and decays of the triplet-like Higgs bosons are essentially determined by inputting four free parameters i.e., $v_\Delta$, $m_L$, $\Delta m^2$ and $\lambda_4$, where 
the last one determines the mixing angle $\alpha$.

%%%%%%%%%%%%%%%%%%%%%%%%%%%%%%%%%%%%%%%%%%%%%%%%%%
%\section{Predictions for $m_W$ and $\Delta \rho$ \label{sec:wmass}}
\noindent
{\it One-loop corrected EW parameters---}
%%%%%%%%%%%%%%%%%%%%%%%%%%%%%%%%%%%%%%%%%%%%%%%%%%
We calculate one-loop corrected $m_W$, $\rho$, $s_W^{}$ and the decay width of the Z boson based on the on-shell renormalization scheme developed in Refs.~\cite{Kanemura:2012rs,Aoki:2012jj}. 
We particularly use the so-called ``Scheme-II'' defined in~\cite{Aoki:2012jj}, where $\alpha_{\rm em}$, $G_F$, $m_Z$ and $v_\Delta$ are chosen as the EW input parameters. 

In order to calculate these EW parameters, we evaluate the $\Delta r$ parameter which modifies the tree level relation of the EW parameters as 
\begin{align}
G_F = \frac{\pi\alpha_{\rm em}}{\sqrt{2}m_W^2s_W^2(1 - \Delta r)}, 
\end{align}
where 
\begin{align}
\Delta r &= \Delta\alpha_{\text{em}}-\frac{c_W^2}{s_W^2}\Delta\rho_{\rm loop}+\Delta r_{\text{rem}}  \label{delr3}. 
\end{align}
In the above expression, 
$\Delta \alpha_{\text{em}}$, $\Delta\rho_{\rm loop}$ and $\Delta r_{\text{rem}}$ respectively denotes the shift of the fine structure constant, the one-loop correction to the rho parameter and the remaining part of $\Delta r$. 
Each contribution can be expressed as 
\begin{align}
\Delta \alpha_{\text{em}} &= \text{Re}\left[\Pi_{\gamma\gamma}'(0)-\Pi_{\gamma\gamma}^\prime(m_Z^2) \right],\label{del_alpha}\\
\Delta \rho_{\rm loop}  &= \text{Re}\left[\frac{\Pi_{ZZ}(0)}{m_Z^2}-\frac{\Pi_{WW}(0)}{m_W^2} + \frac{2s_W}{c_W}\frac{\Pi_{Z\gamma }(0)}{m_Z^2}-\frac{\sin2\beta'}{1 + \cos^2\beta'}\delta \beta'\right],\label{del_rho}\\
\Delta r_{\text{rem}}&=\frac{c_W^2}{s_W^2}\text{Re}\left[\frac{\Pi_{ZZ}(0)}{m_Z^2}-\frac{\Pi_{ZZ}(m_Z^2)}{m_Z^2}\right]+
\left(1-\frac{c_W^2}{s_W^2}\right)\text{Re}\left[\frac{\Pi_{WW}(0)}{m_W^2}-\frac{\Pi_{WW}(m_W^2)}{m_W^2}\right]\notag\\
&+\text{Re}\Pi_{\gamma\gamma}'(m_Z^2)+\delta_{VB}, \label{del_rem}
\end{align}
where $\Pi_{VV'}$ are the contribution from 1PI diagrams for the transverse part of gauge boson two point functions, and $\delta_{VB}$ is the vertex and box corrections to light fermion scatterings $f\bar{f} \to f\bar{f}$. 
See Refs.~\cite{Kanemura:2012rs,Aoki:2012jj} for their concrete analytic expressions. 
In the expression for the one-loop correction to the rho parameter $\Delta \rho_{\rm loop}$, 
$\delta \beta'$ denotes the counter term for the $A$-$G^0$ mixing $\beta'$, see Eq.~(\ref{mixing1}). This can be determined by imposing the on-shell condition for the $A$-$G^0$ mixing~\cite{Kanemura:2012rs}: 
\begin{align}
\hat{\Pi}_{AG^0}(0)=\hat{\Pi}_{AG^0}(m_A^2) = 0, 
\end{align}
where $\hat{\Pi}_{AG^0}$ is the renormalized two point function of the $A$-$G^0$ mixing\footnote{These two conditions determine $\delta \beta'$ and the off-diagonal wave function renormalization factor $\delta C_{AG^0}$ defined in Ref.~\cite{Aoki:2012jj}, where the latter is not relevant to the analysis done in this Letter.  }. 

The one-loop corrected W boson mass and the weak mixing angle are then calculated  by the on-shell scheme~\cite{Sirlin:1980nh,Aoki:2012jj} as
\begin{align}
(m_W^2)_{\text{ren}} &= 
\frac{m_Z^2(1 + \cos^2\beta')}{4}\left[1+\sqrt{1-\frac{8}{1 + \cos^2\beta'}\frac{\pi\alpha_{\text{em}}}{\sqrt{2}G_Fm_Z^2(1-\Delta r)}}\right], \\
(s_W^2)_{\text{ren}} &= 
\frac{1 + \cos^2\beta'}{4}\left[1-\sqrt{1-\frac{8}{1 + \cos^2\beta'}\frac{\pi\alpha_{\text{em}}}{\sqrt{2}G_Fm_Z^2(1-\Delta r)}}\right].
\end{align}
In addition, the one-loop corrected rho parameter is given by $\rho_{\rm ren}= \rho_{\rm tree} + \Delta\rho_{\rm loop}$. 
It has been known that the effective weak mixing angle, denoting as $\theta_{\rm eff}^f$, has currently most precisely been measured from 
the $Z$ pole observables, which is defined by 
\begin{align} 
\sin^2 \theta_{\rm eff}^f = \frac{1}{4|Q_f|}\left[1 - \text{Re}\left(\frac{g_V^f}{g_A^f}\right)\right]_{p^2 = m_Z^2}, 
\end{align}
where $g_V^f$ and $g_A^f$ are respectively the vector and axial vector part of the effective $Zf\bar{f}$ vertex. They are defined as~\cite{ALEPH:2005ab}  
\begin{align}
g_V^f &= \sqrt{\rho_{\rm ren}}\left[I_f -2Q_f (s_W^2)_{\text{ren}}\left(1 + \frac{(c_W^{})_{\text{ren}}}{(s_W^{})_{\text{ren}}}\frac{\hat{\Pi}_{Z\gamma}(m_Z^2)}{m_Z^2}\right)\right], \\
g_A^f &= \sqrt{\rho_{\rm ren}}I_f,
\end{align}
with $\hat{\Pi}_{Z\gamma}$ being the renormalized $Z$--$\gamma$ mixing function and $I_f$ being the third component of the isospin for a fermion $f$. 
We find 
\begin{align}
\sin^2 \theta_{\rm eff}^f = s_W^2 - \frac{c_W^2s_W^2}{c_W^2 - s_W^2}\Delta \rho_{\rm loop} + s_W^{}c_W^{}\Delta r_{\rm rem}', 
\end{align}
where 
\begin{align}
\Delta r_{\rm rem}' &= \frac{\Pi_{Z\gamma}(m_Z^2) - \Pi_{Z\gamma}(0)}{m_Z^2} + \frac{c_W^{}}{s_W^{}}
\left[\frac{\Pi_{ZZ}(m_Z^2) - \Pi_{ZZ}(0)}{m_Z^2} - \frac{\Pi_{WW}(m_W^2) - \Pi_{WW}(0)}{m_W^2}\right]. 
%& = \frac{g^2}{16\pi}\left(\frac{c_W}{s_W}U - \frac{s_W}{c_W}S \right). 
\end{align}
By using the effective couplings $g_V^f$ and $g_A^f$, the EW radiatively corrected decay rate of the Z boson are calculated as 
\begin{align}
\Gamma(Z \to f\bar{f}) = \frac{\sqrt{2}m_Z^2G_F}{12\pi}[(g_V^f)^2 + (g_A^f)^2]. 
\end{align}

%%%%%%%%%%%%%%%%%%%%%%%%%%%%%%%%%%%%%%%%%%%%%%%%%%
%\section{Predictions for $m_W$ and $\Delta \rho$ \label{sec:wmass}}
\noindent
{\it Results---}
%%%%%%%%%%%%%%%%%%%%%%%%%%%%%%%%%%%%%%%%%%%%%%%%%%
We use the following input values for the SM parameters~\cite{Zyla:2020zbs}: 
\begin{align}
&\alpha_{\rm em}^{-1} = 137.036,~~G_F = 1.1663787\times 10^{-5}~\text{GeV}^{-2},~~m_Z=91.1876~\text{GeV}, \notag\\
&m_h=125.25~\text{GeV},~~m_t = 173.1~\text{GeV}. 
\end{align}
In the following numerical analysis, we fix $\lambda_4 = 0$. The treatment of the other input parameters $\Delta m^2$, $m_L$ and $v_\Delta$ is described as follows, where 
the size of the squared mass difference $\Delta m^2$ is constrained by the perturbative unitarity bound~\cite{Aoki:2007ah,Arhrib:2011uy} which is taken into account in the analysis. 
In order to constrain the parameter space in our model, we introduce 
\begin{align}
&\Delta m_W \equiv (m_W)_{\rm ren} - (m_W)_{\rm ren}|_{\rm ref}, \quad 
\Delta \rho \equiv \rho_{\rm ren} - \rho_{\rm ren}|_{\rm ref}, \notag\\
&\Delta s_{\rm eff}^2 \equiv \sin^2\theta_{\rm eff}^{\ell} - \sin^2\theta_{\rm eff}^{\ell}|_{\rm ref}, \quad 
\Delta \Gamma_{\rm lep} \equiv \Gamma(Z \to \ell\ell) - \Gamma(Z \to \ell\ell)_{\rm ref}~~~(\ell = e,\mu),  \label{eq:delta}
\end{align}
where $X|_{\rm ref}$ are the reference values of the quantity $X$, which are obtained by taking the decoupling limit $v_\Delta \to 0$, $\Delta m^2 \to 0$ and $m_L \to \infty$. 
We then require that new physics contributions to $\rho_{\rm ren}$, $\sin^2\theta_{\rm eff}^{\ell}$ and $\Gamma(Z \to \ell\ell)$ are given within the $2\sigma$ error of each measurement assuming 
the SM predictions being their central values~\cite{Zyla:2020zbs}: 
\begin{align}
|\Delta \rho| \leq 1.8\times 10^{-3},~~|\Delta s_{\rm eff}^2| \leq 6.6\times 10^{-4},~~|\Delta \Gamma_{\rm lep}|\leq 0.17~\text{MeV}.   \label{eq:const1}
\end{align}
For $m_W$, we use the measured value at CDF II, i.e., $m_W^{\text{CDF II}} = 80.4335 \pm 0.0094$ GeV and the central value of the SM prediction $m_W^{\rm SM} = 80.357$ GeV. 
We then impose the constraint on the W mass at $2\sigma$ level as 
\begin{align}
57.7~\text{MeV} \leq \Delta m_W \leq 95.3~\text{MeV}. \label{eq:const2}
\end{align}

\begin{figure}[!t]
\begin{center}
 \includegraphics[width=55mm]{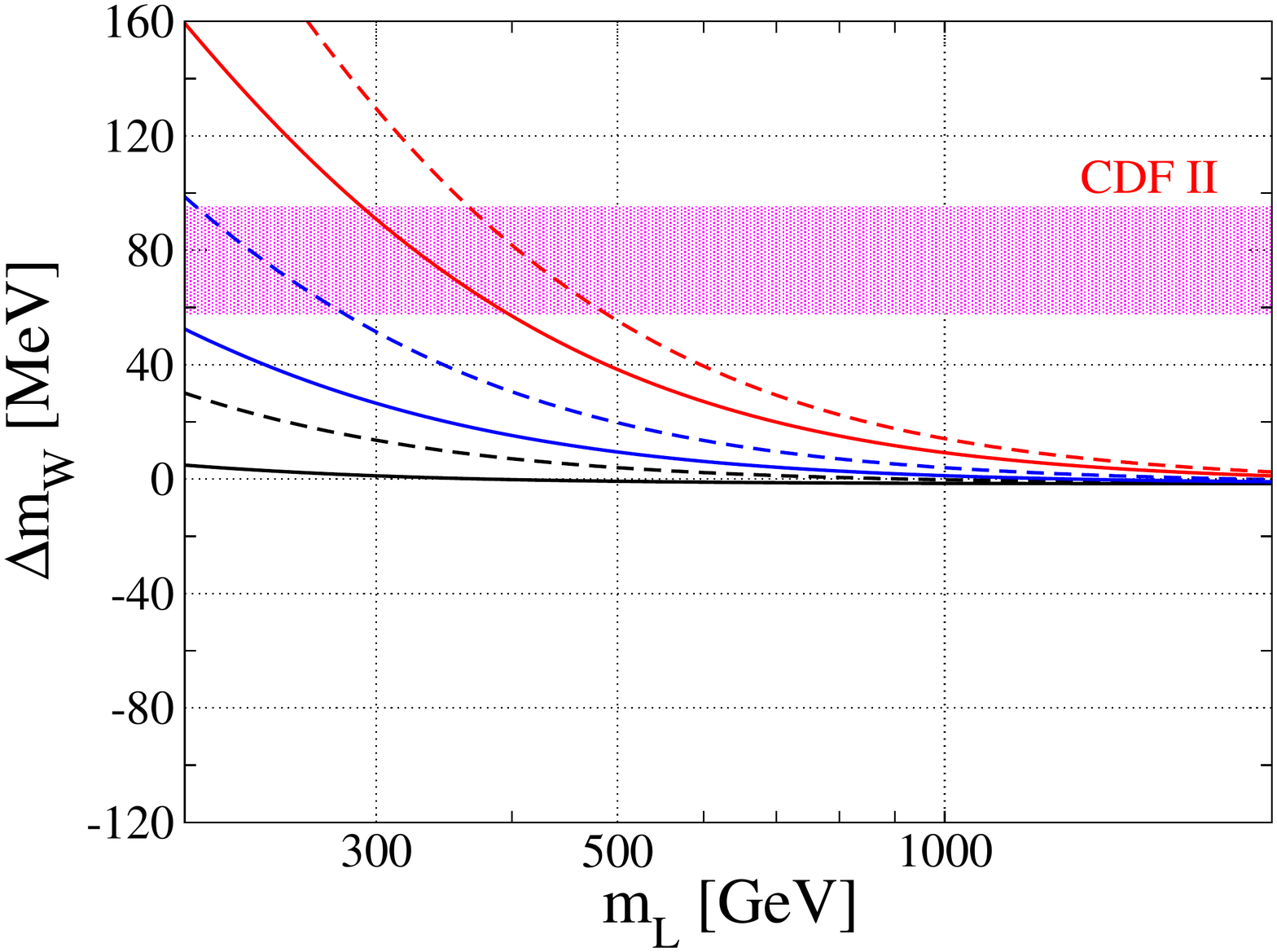}\hspace{-5mm}
 \includegraphics[width=55mm]{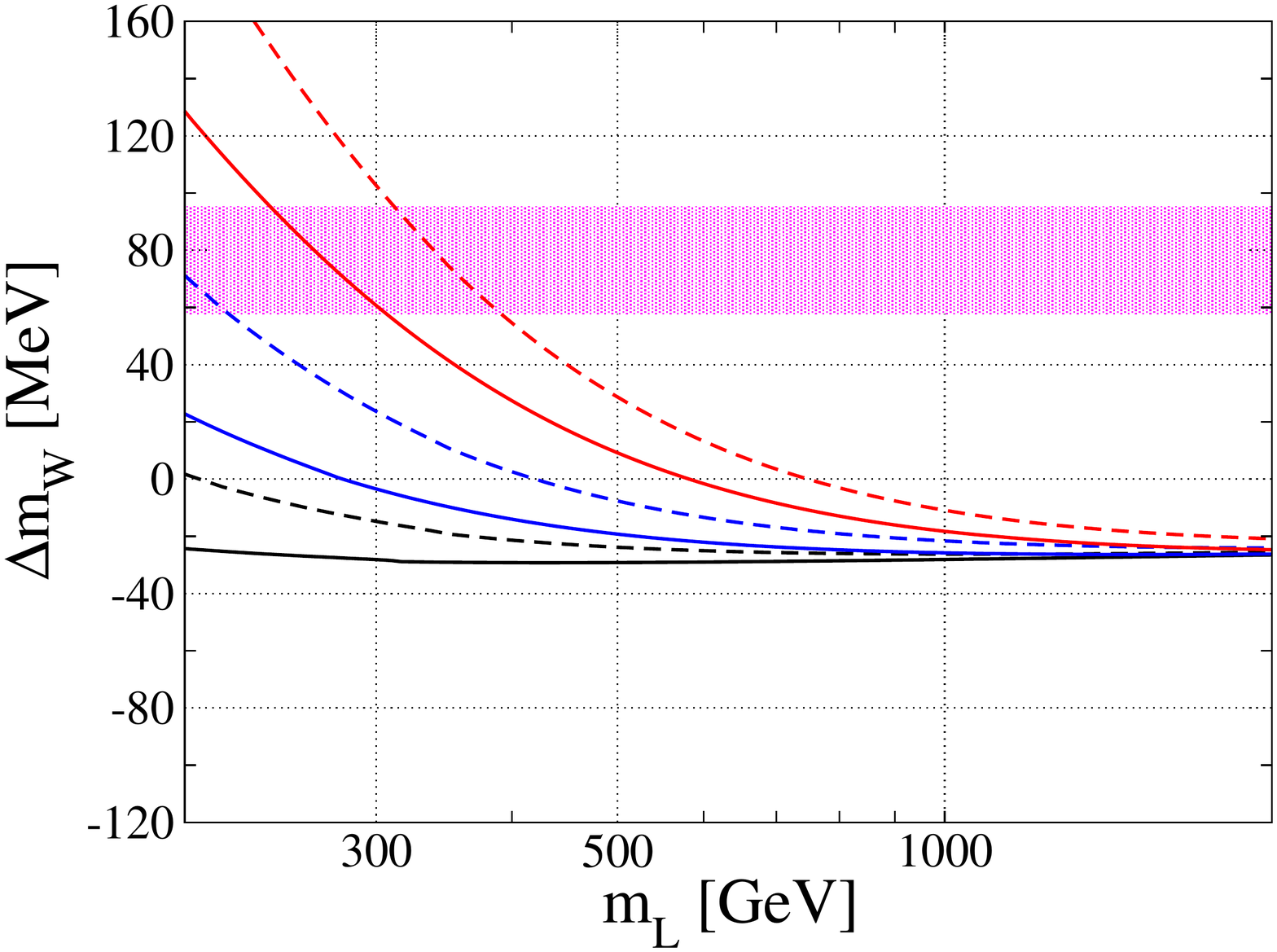}\hspace{-5mm}
 \includegraphics[width=55mm]{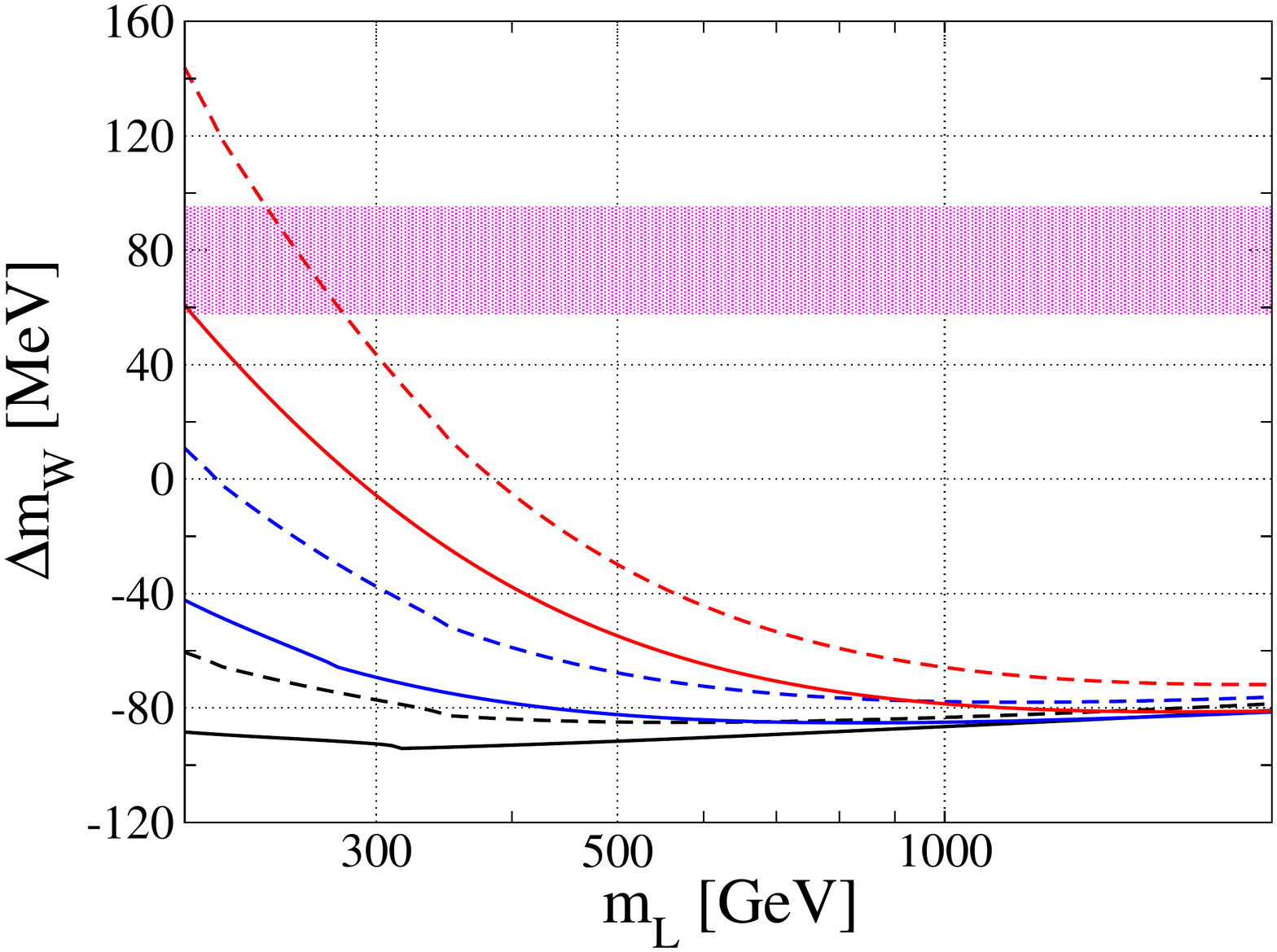}\\ \vspace{-2mm}
 \includegraphics[width=55mm]{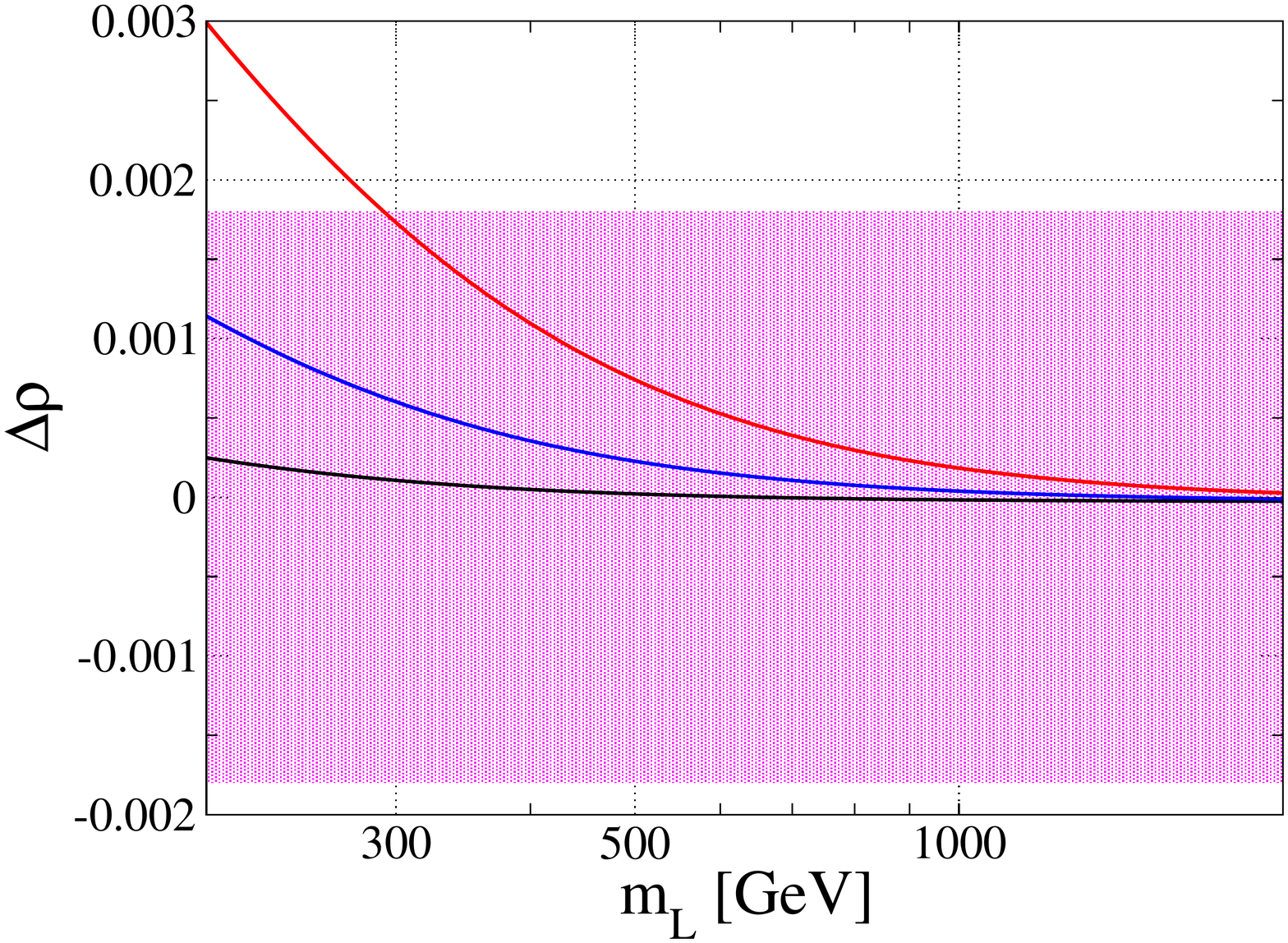}\hspace{-5mm}
 \includegraphics[width=55mm]{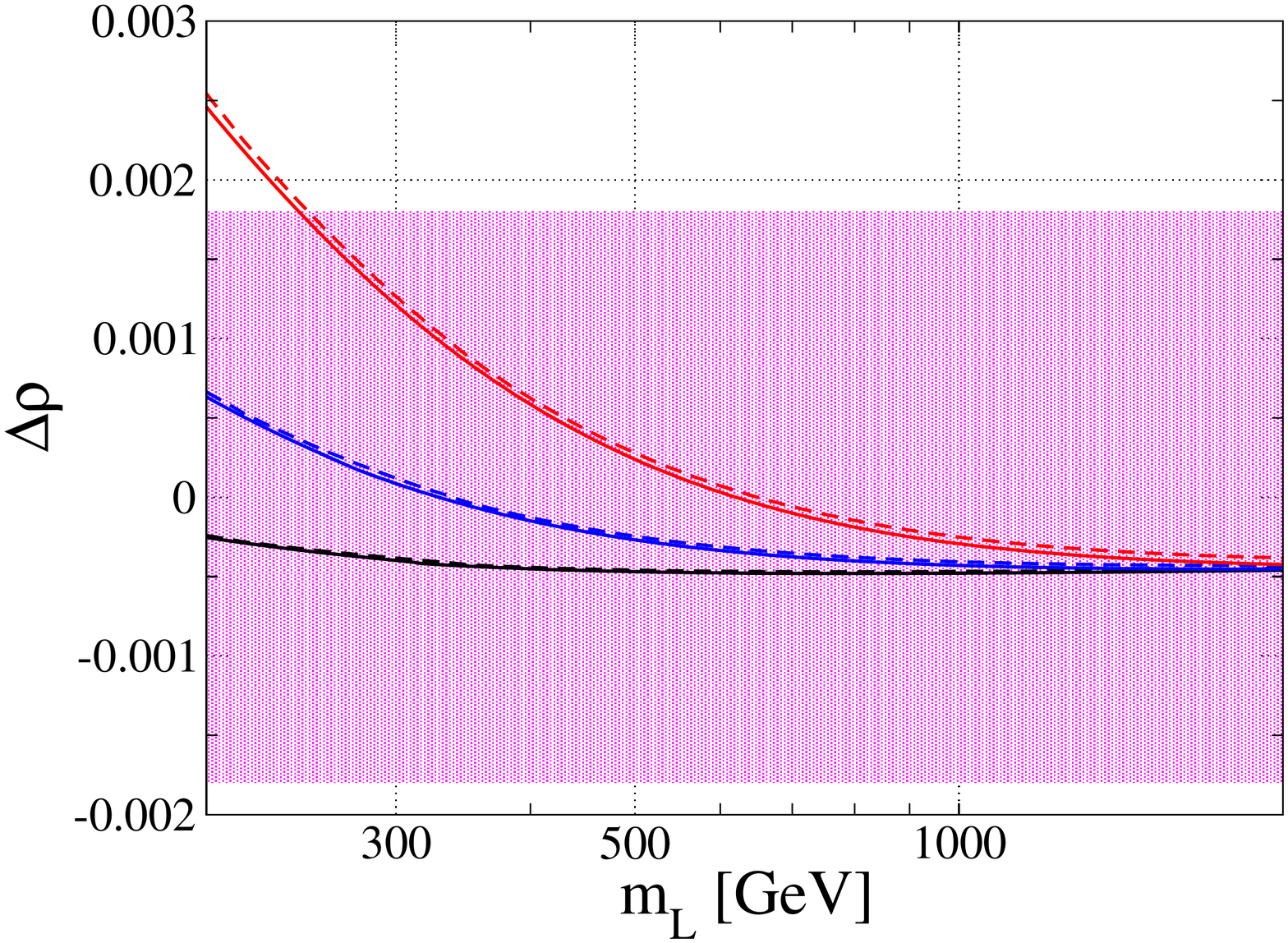}\hspace{-5mm}
 \includegraphics[width=55mm]{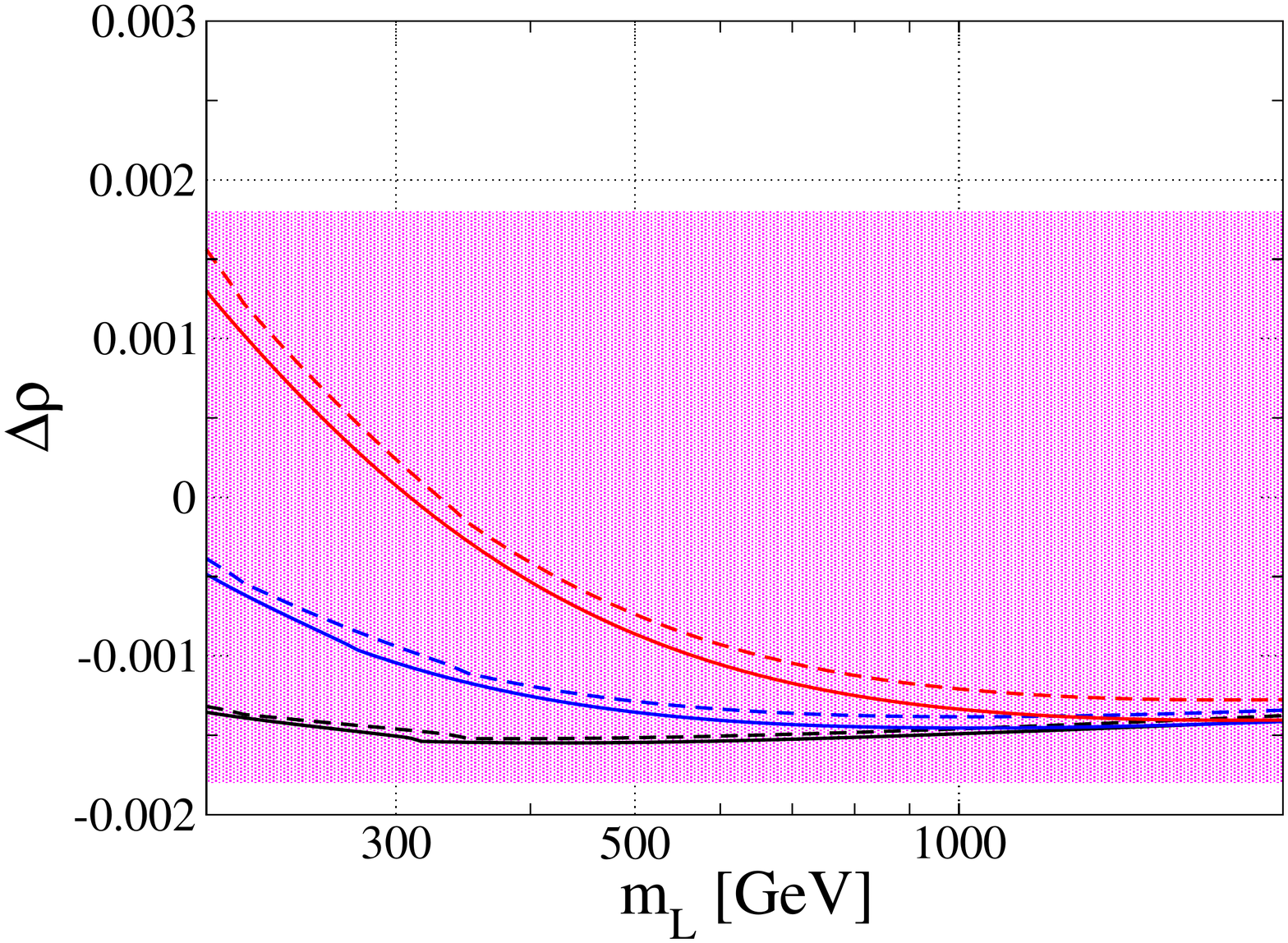}\\\vspace{-2mm}
 \includegraphics[width=55mm]{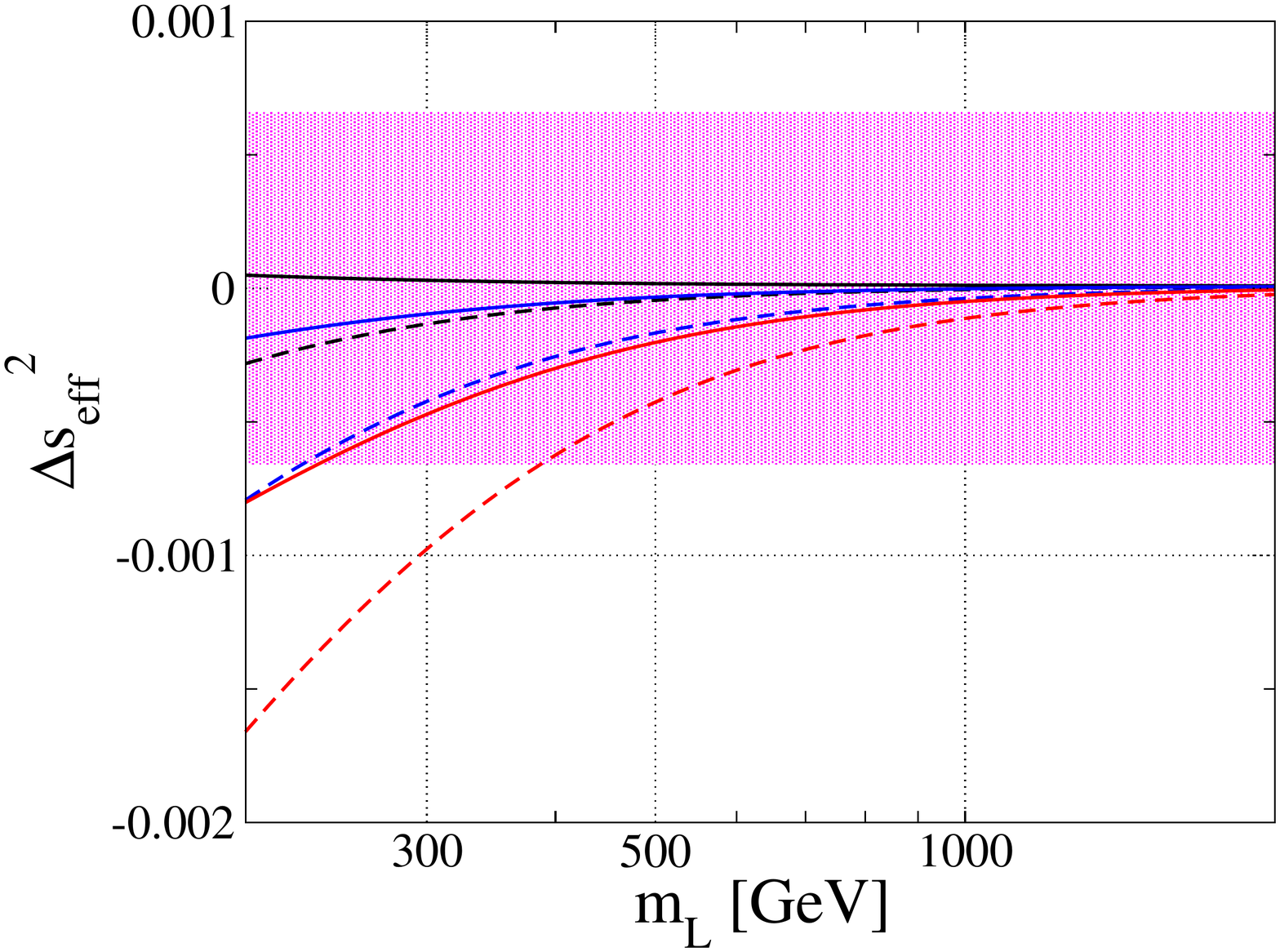}\hspace{-5mm}
 \includegraphics[width=55mm]{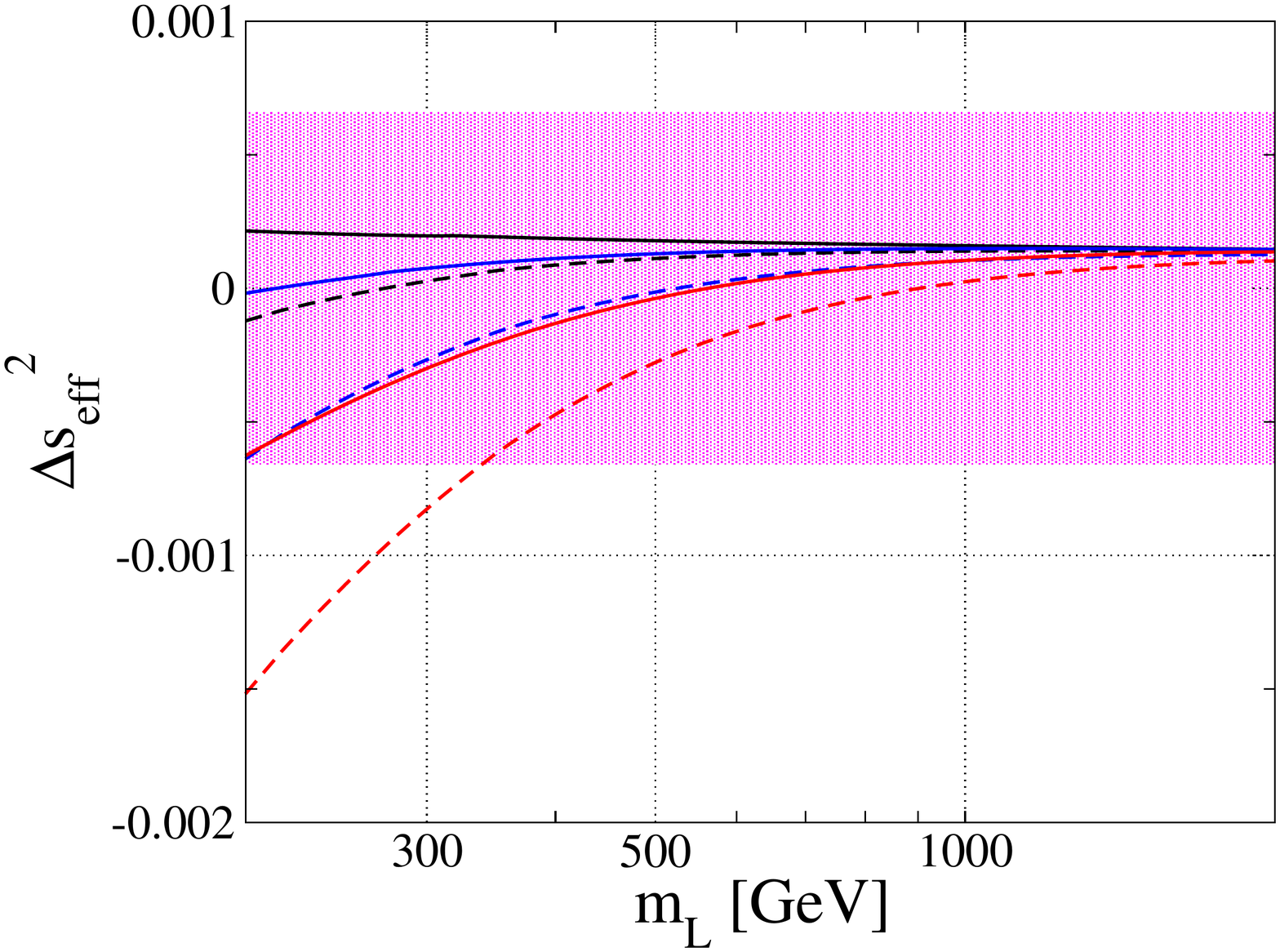}\hspace{-5mm}
 \includegraphics[width=55mm]{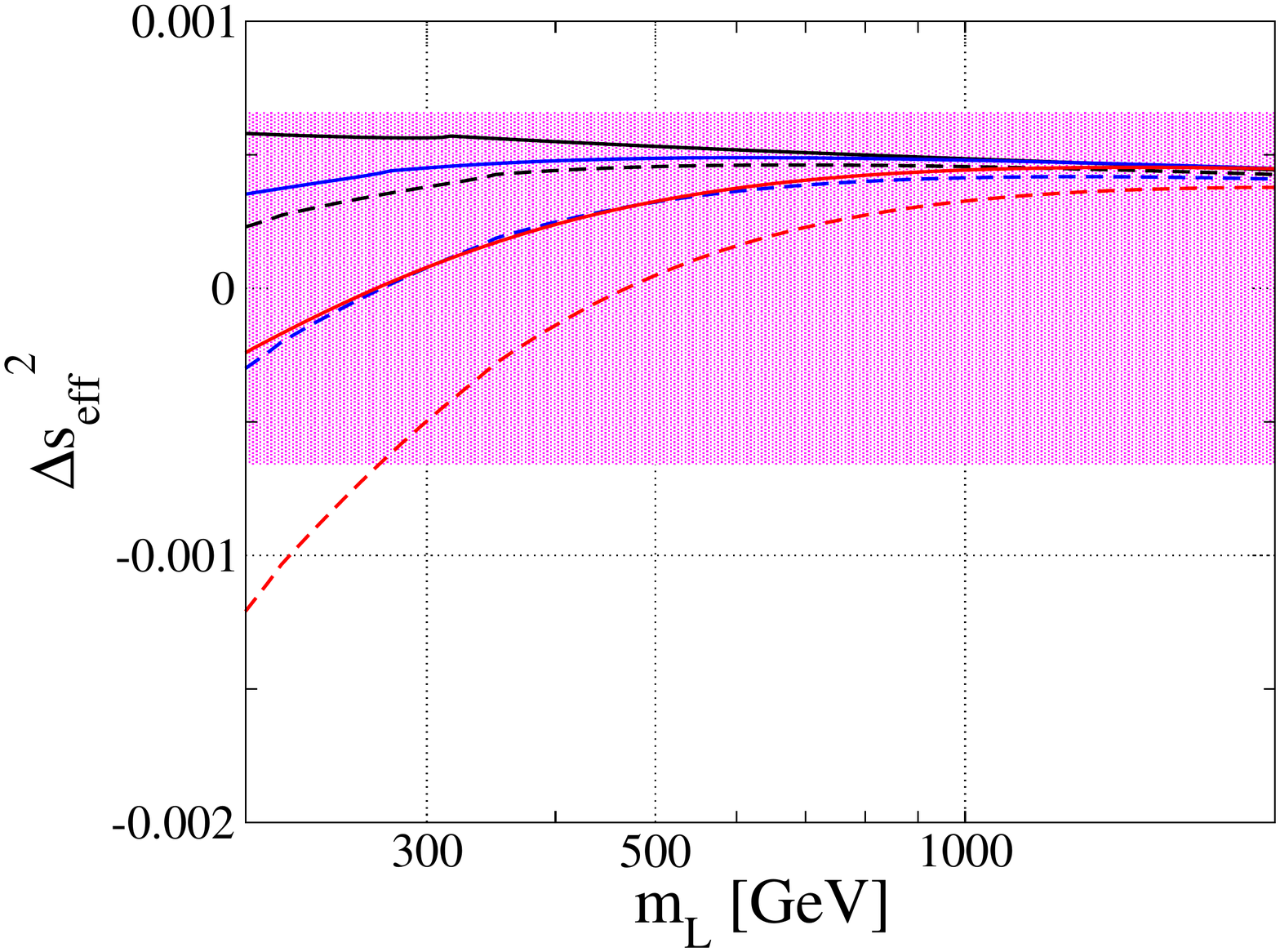}\\\vspace{-2mm}
 \includegraphics[width=55mm]{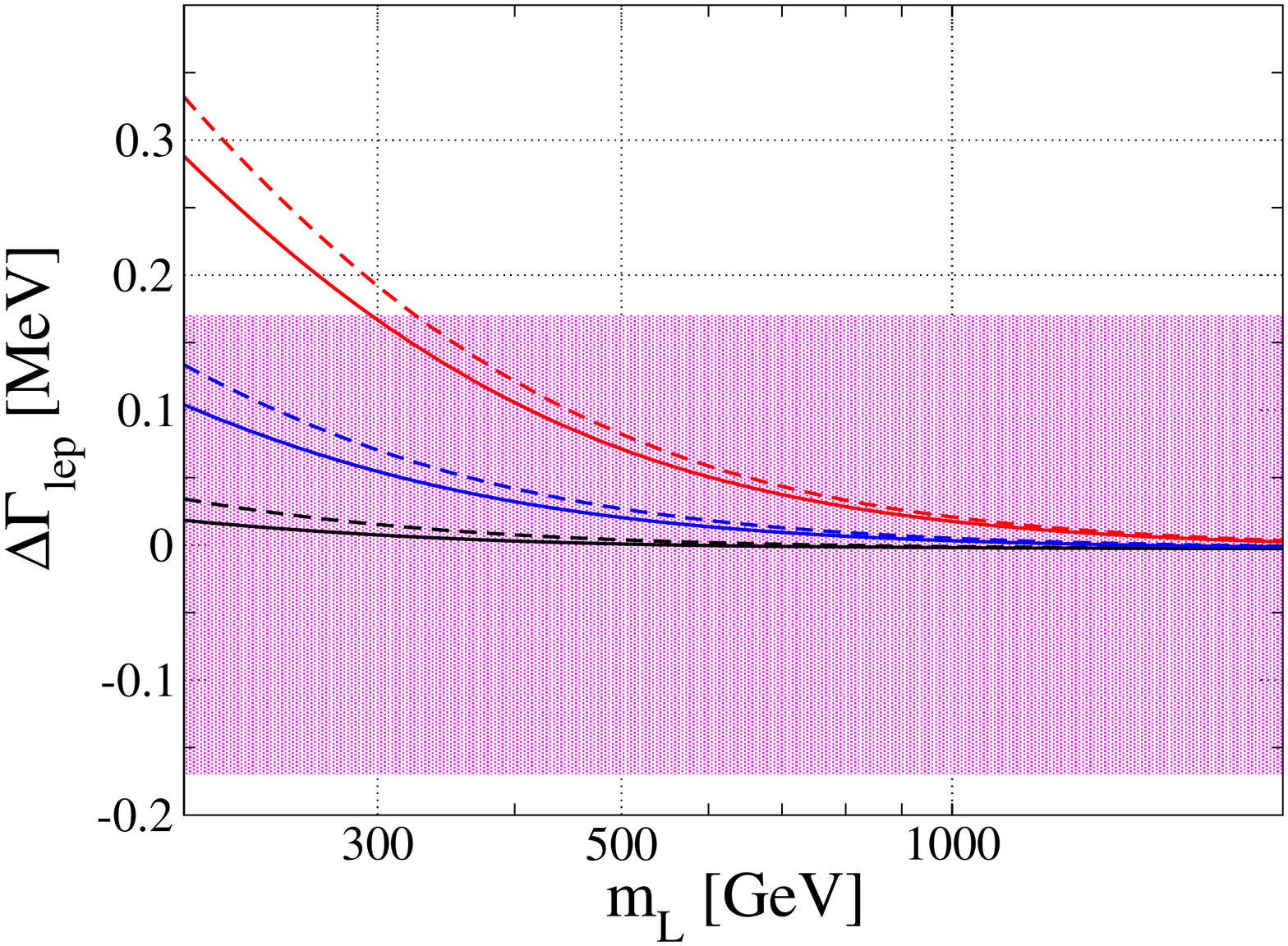}\hspace{-5mm}
 \includegraphics[width=55mm]{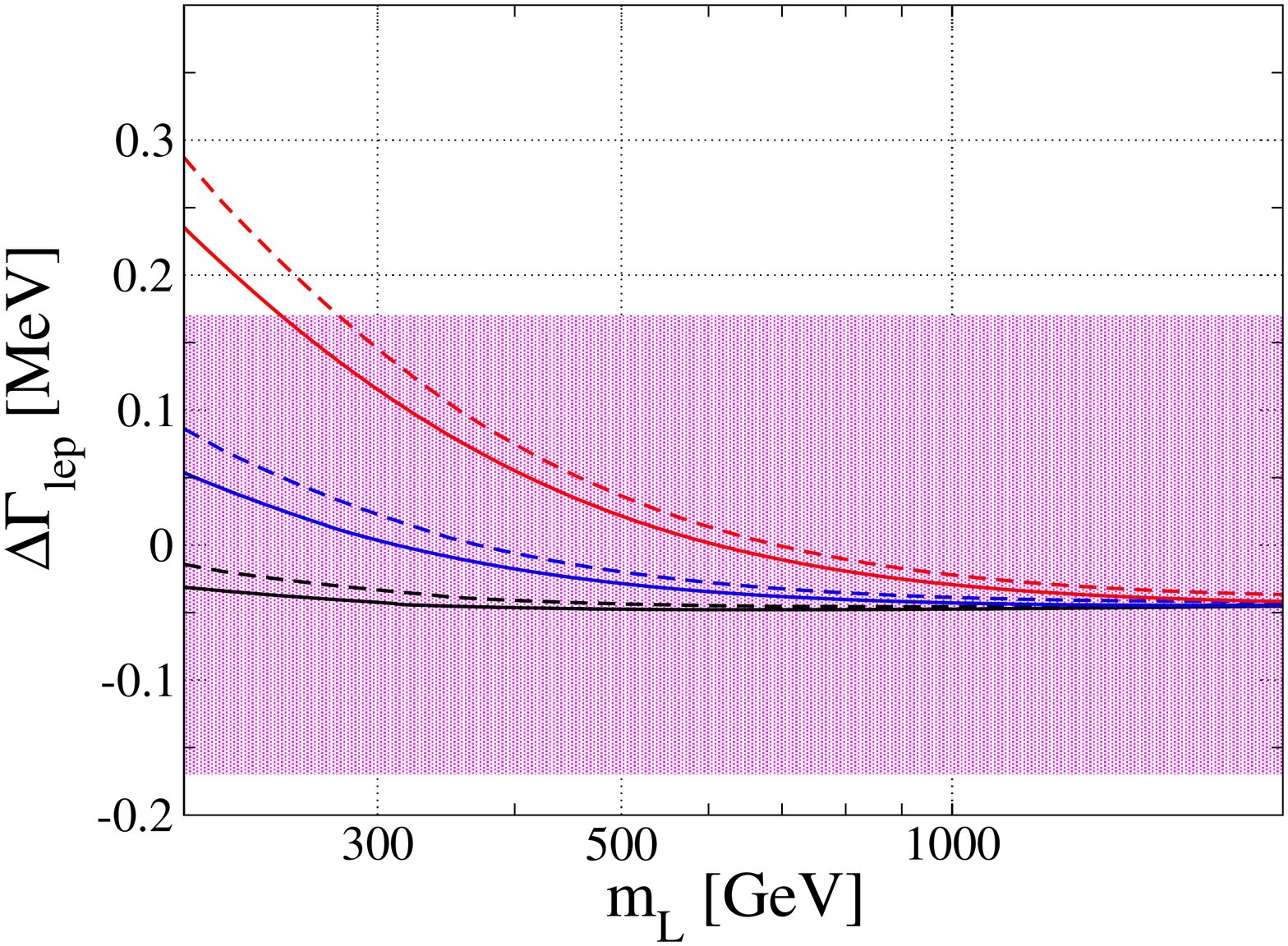}\hspace{-5mm}
 \includegraphics[width=55mm]{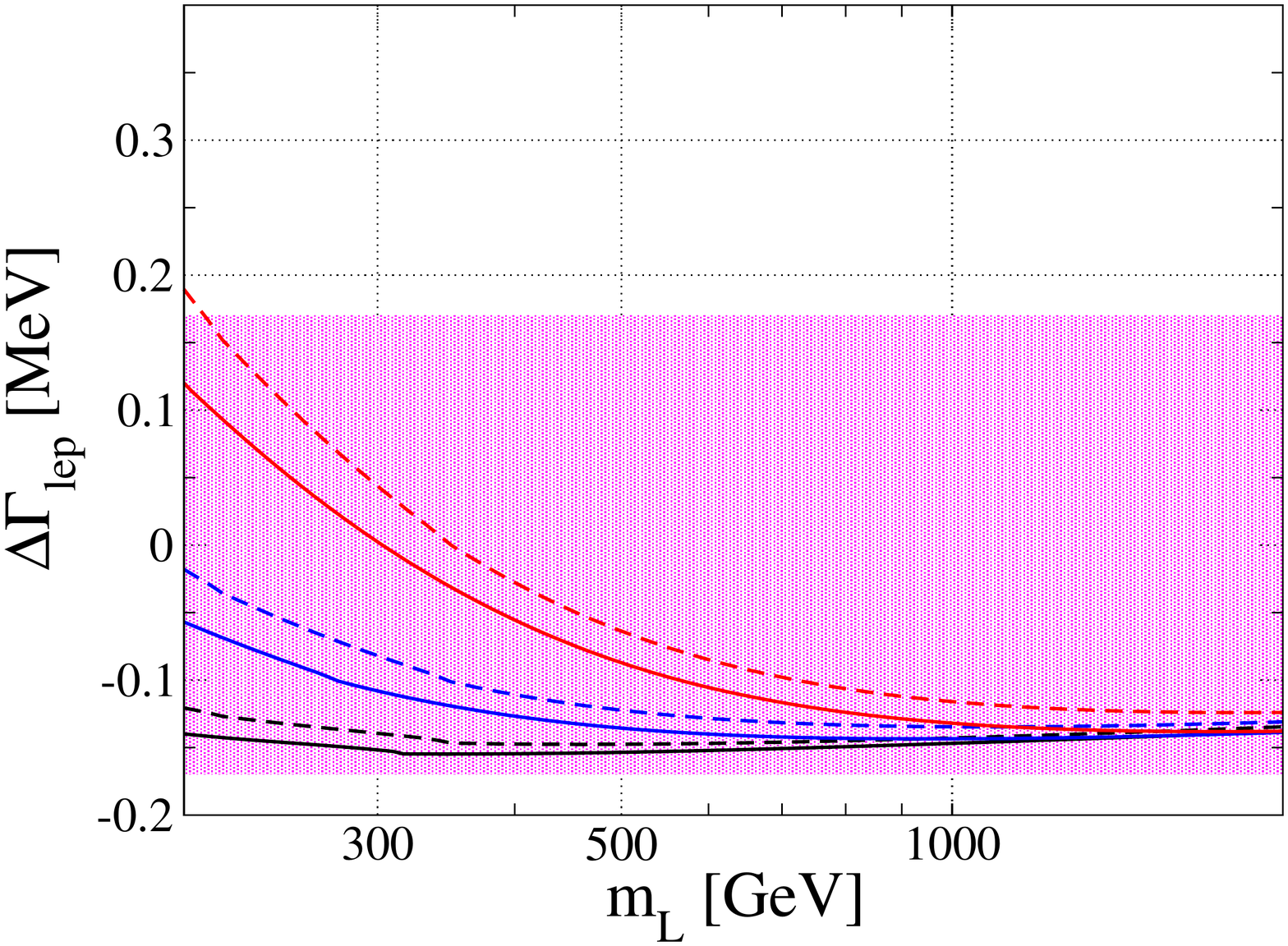}
   \caption{One-loop predictions of $\Delta m_W$, $\Delta \rho$, $\Delta s_{\rm eff}^2$ and $\Delta\Gamma_{\rm lep}$ defined in Eq.~(\ref{eq:delta}) as a function of the lightest triplet-like Higgs boson mass $m_L$. 
We take $v_\Delta$ to be 1 GeV, 4 GeV and 7 GeV for the left, center and right panels, respectively.
The solid (dashed) curves show the case with $\Delta m^2 < 0$ ($\Delta m^2 > 0$ ), while the 
black, blue and red curves denote the case with $|\Delta m^2| = 100^2$, $150^2$ and $200^2$ GeV$^2$, respectively. 
The magenta band shows the 2$\sigma$ region allowed by the experiments. }
   \label{fig:1}
\end{center}
\end{figure}

In Fig.~\ref{fig:1}, we show the predictions of $\Delta m_W$, $\Delta \rho$, $\Delta s_{\rm eff}^2$ and $\Delta \Gamma_{\rm lep}$ from the top to bottom as a function of the mass of the lightest triplet-like Higgs boson $m_L$. 
We take $v_\Delta$ to be 1 GeV (left), 4 GeV (center) and 7 GeV (right), where the case with $v_\Delta < 1$ GeV is almost the same as that with $v_\Delta = 1$ GeV. 
The magenta region represents the one given in Eqs.~(\ref{eq:const1}) and (\ref{eq:const2}). 
Clearly, all these predictions approach to the SM values at large $m_L$ for the case with small $v_\Delta$ as expected by the decoupling theorem. 
On the other hand, for larger $v_\Delta$, these predictions can deviate from the SM values even at the large mass region, because its tree level effects on these observables can be significant. 
It is seen that the case with larger $|\Delta m^2|$ the values of $\Delta m_W$, $\Delta \rho$ and $\Delta \Gamma_{\rm lep}$ take larger values, while $\Delta s_{\rm eff}^2$ becomes smaller. 
In addition, we see that the case with $\Delta m^2 > 0$ (dashed curves) shows larger deviations of the EW parameters as compared with the case for $\Delta m^2 < 0$ (solid curves). %
We find that the measured W boson mass at CDF II (magenta region) can be explained when $m_L$ is taken to be 200-500 GeV with $\sqrt{|\Delta m^2|}$ to be of order 100 GeV. 

\begin{figure}[!t]
\begin{center}
 \includegraphics[width=120mm]{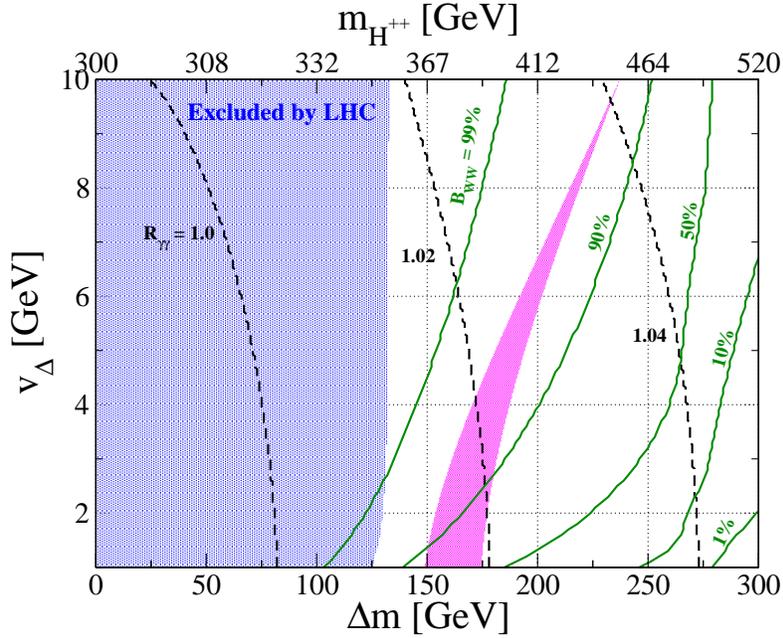}
   \caption{Results for the case with $m_L = 300$ GeV and $\Delta m^2 > 0$, i.e., $H^{\pm\pm}$ are the heaviest, on the $\sqrt{\Delta m^2}$-$v_\Delta$ plane. 
We also show the corresponding value of $m_{H^{\pm\pm}}$ on the upper side of the $x$-axis. 
Regions shaded in magenta (blue) satisfy Eqs.~(\ref{eq:const1}) and (\ref{eq:const2}) (excluded by the direct search for $H^{\pm\pm} \to W^\pm W^\pm$ at LHC). 
The solid and dashed curves show the contour of $B_{WW}\equiv {\cal B}(H^{\pm\pm} \to W^\pm W^\pm)$ and $R_{\gamma\gamma}\equiv \Gamma_{h \to \gamma\gamma}|_{\rm {HTM}}/\Gamma_{h \to \gamma\gamma}|_{\rm {SM}}$, respectively. }
   \label{fig:2}
\end{center}
\end{figure}

In order to clarify whether the parameter set reproducing the measured value of $m_W$ at CDF II is allowed by all the constraints given in Eq.~(\ref{eq:const1}), we show Fig.~\ref{fig:2}, in which 
the region shaded in magenta satisfies the condition given in Eqs.~(\ref{eq:const1}) and (\ref{eq:const2}) in the case of $m_{L} = 300$ GeV and $\Delta m^2 > 0$. 
In our scenario with $v_\Delta > 1$ GeV, $H^{\pm\pm}$ mainly decay into the same sign diboson $W^\pm W^\pm$~\cite{Kanemura:2014ipa,Kanemura:2014goa,Kanemura:2013vxa,Chiang:2012dk} 
or $H^\pm W^{\pm (*)}$, see e.g., Refs.~\cite{Melfo:2011nx,Aoki:2011pz} for the LHC phenomenology with $H^{\pm\pm} \to H^\pm W^{\pm (*)}$. 
When $H^{\pm\pm}$ decay into the diboson with hundred percent, the current lower limit on $m_{H^{\pm\pm}}$ has been taken to be about 350 GeV at LHC~\cite{ATLAS:2021jol}. 
This limit can be relaxed when the branching ratio ${\cal B}(H^{\pm\pm} \to W^\pm W^\pm)$ decreases due to the cascade decay mode. 
We show the region excluded by the direct search at LHC limit from the LHC data  (shaded in blue). 
It is now clear that the region satisfying Eqs.~(\ref{eq:const1}) and (\ref{eq:const2}) is also allowed by the direct search. 
We also display the deviation in the decay rate of the $h\to \gamma\gamma$ mode~\cite{Shifman:1979eb,Ellis:1975ap}, 
which is shown by the dashed contour for $R_{\gamma\gamma}$ defined as 
\begin{align}
R_{\gamma\gamma} \equiv \frac{\Gamma_{h \to \gamma\gamma}|_{\rm {HTM}}}{\Gamma_{h \to \gamma\gamma}|_{\rm {SM}}}. 
\end{align}
It is seen that the deviation is given to be a few percent level which is well inside the current bound at LHC~\cite{ATLAS:2021vrm,CMS:2020gsy}.

%%%%%%%%%%%%%%%%%%%%%%%%%%%%%%%%%%%%%%%%%%%%%%%%%%

%%%%%%%%%%%%%%%%%%%%%%%%%%%%%%%%%%%%%%%%%%%%%%%%%%
%%%%%%%%%%%%%%%%%%%%%%%%%%%%%%%%%%%%%%%%%%%%%%%%%%
%\section{Conclusions \label{sec:conclusions}} 
\noindent
{\it Conclusions ---}
%%%%%%%%%%%%%%%%%%%%%%%%%%%%%%%%%%%%%%%%%%%%%%%%%%
We have discussed the implication of the large discrepancy between the W boson mass measured at CDF II and its SM prediction in the HTM. 
In this model, the W boson mass can significantly be modified from the SM prediction by both the effects from the triplet VEV $v_\Delta$ at tree level 
and radiative corrections of triplet-like Higgs boson loops. 
We have found that the anomaly in the W boson mass can be explained under the constraint from the EW rho parameter, the effective weak mixing angle and the partial width of the leptonic $Z$ decay when 
there is a non-zero mass splitting among $H^{\pm\pm}$, $H^{\pm}$, $A$ and $H$ with $v_\Delta$ of order GeV. 
For instance, when we take $H/A$ are the lightest triplet-like Higgs bosons ($\Delta m^2 > 0$) with their mass of 300 GeV, 
the anomaly can be explained by taking $\sqrt{\Delta m^2} \simeq 160$ GeV (200 GeV) for $v_\Delta = 2$ GeV (6 GeV). 
In this scenario, $H^{\pm\pm}$ can mainly decay into the same sign diboson or $H^\pm W^\mp$ depending on the size of the mass splitting. 
We have confirmed that the region compatible with the CDF anomaly is allowed by the current searches for $H^{\pm\pm}$ decaying into diboson at LHC. 
In addition, we have seen that the decay rate of the $h \to \gamma\gamma$ deviates from the SM prediction with a few percent level, which is well 
inside the current measured value at LHC. 

\vspace*{4mm}

\begin{acknowledgments}
This work is supported in part by the JSPS KAKENHI Grant No. 20H00160 [S.K.] and Early-Career Scientists No. 19K14714 [K.Y.].
\end{acknowledgments}

\bibliography{references}

\end{document}